\documentclass[aps,prd,twocolumn,nofootinbib]{revtex4-2}

\usepackage[colorlinks]{hyperref}
\usepackage{amssymb,amsmath}
\usepackage{graphicx}
\usepackage{subcaption}

\begin{document}
\title{Manifestation of antiquark nuggets in collisions with the Earth}
\author{V. V. Flambaum}\email{v.flambaum@unsw.edu.au}
\author{I. B. Samsonov}\email{igor.samsonov@unsw.edu.au}
\author{G. K. Vong}\email{g.vong@unsw.edu.au}
\address{School of Physics, University of New South Wales, Sydney 2052, Australia}

\begin{abstract}
Antiquark nuggets are hypothetical compact composite objects conjectured to account for a significant fraction of dark matter in the Universe. In contrast to quark nuggets, these objects consist of antimatter. They may remain undetected if they possess a sufficiently small cross section relative to their mass. In this paper, we investigate the allowed region in the parameter space of this model that is consistent with the observed neutrino flux from the Sun and the Earth, and the nonobservation of seismic events with specific signatures of dark matter particles. We found the allowed values of the antibaryon charge number in this model to be in the interval $2\times 10^{24}<A<8\times 10^{25}$, while the probability of nucleon  annihilation upon collisions with the antiquark core is constrained by $0.1\lesssim \kappa<0.25$. These values of $A$ and $\kappa$ are, however, constrained by the IceCube experiment and nonobservation of impacts of antiquark nuggets on humans. Although very large values of the antibaryon charge, $A>10^{33}$, are not fully excluded by the present study, we show that they conflict with the nonobservation of rare catastrophic explosionlike events on the Earth.
\end{abstract}

\maketitle

\section{Introduction}

Despite long-term efforts, the detection of dark matter (DM) particles remains an elusive problem in physics, with little progress in understanding the nature of these particles. Although the Standard Model of elementary particles allows for consistent extensions with various candidate DM particles, such as supersymmetric particles, axions, and sterile neutrinos, none of these candidates has received any experimental evidence. 

An alternative approach to this problem is to assume no new particles beyond the Standard Model and look for compact composite objects that may have feeble interactions with visible matter and remain undiscovered due to a small interaction cross section relative to their mass. The most notable example of such a model is strangelets, hypothetical droplets of quark matter stabilized by strange quarks \cite{Witten84,Farhi,Glashow}. More generally, macroscopic composite dark matter particles were studied in a series of works \cite{Starkman1,Starkman2,Starkman3,Starkman4,Starkman5,Starkman6,Starkman7,Starkman8,Starkman9,Starkman10,Caloni}.

In this paper, we consider an extension of the strangelet model developed by A. Zhitnitsky and his collaborators in a series of papers \cite{DW1,DW2,DW3,Zhitnitsky_2003,CCO}, dubbed axion-quark nuggets (AQNs). Unlike strangelets, this model assumes an axion-pion domain wall that creates pressure on the quark core, providing stability against decay into baryonic matter. Another notable aspect of this model is the assumption that, in addition to quark nuggets (QNs), there are antiquark nuggets that comprise all antimatter, making the total number of quarks and antiquarks in the Universe equal \cite{Zhitnitsky_2003,CCO,Basymmetry,WMAPhaze}.\footnote{See also  earlier proposals for baryon symmetric baryogenesis \cite{Widrow,Sakharov}.}
This idea offers a possible solution to the problem of matter-antimatter asymmetry in the Universe. The present paper investigates the feasibility of this assumption. Therefore, in the remainder of this paper, we will focus on antiquark nuggets (antiQNs) and their possible manifestations.

Let us recall on the structure of antiQN and basic assumptions behind this model. Each antiQN is supposed to have an antiquark core surrounded by a positron cloud needed for electric neutrality and beta-stability of antiQN. AntiQNs are characterized by a mass number $A=|B|$, where $B$ is (negative) baryon charge number, so that the mass may be represented as $m_\text{QN}\simeq A m_p$, with proton mass $m_p$. Unlike ordinary nuclei, the mass number $A$ is supposed to be very large, $A>10^{24}$, because lower values of $A$ are excluded by earlier studies \cite{FS0,FS1,FS2,Garry,Budker2020}, see also Ref.~\cite{Zh-review} for a review. However, this limit was based on some assumptions which will be reconsidered below. 

The density of the antiquark core is supposed to be comparable with or larger than the typical nuclear density. This assumption is effectively taken into account by the relation
\begin{equation}
    R_\text{QN}\simeq 1\,\text{fm}\,A^{1/3}\,,
\end{equation}
where $R_\text{QN}$ is the radius of the antiquark core. Thus, the mass number $A$ is the main parameter in this model which needs to be fixed from various physical (non)manifestations of antiQNs. Note that, in general, $A$ is not a single fixed number, but it is taken from some distribution with a mean $\langle A\rangle$. We will, however, omit the angle brackets for brevity and assume that $A$ denotes this mean mass number of antiQNs. 

The positron cloud around the antiquark core plays important role in this model because it is responsible for thermal radiation from antiQNs when they interact with some medium. In Ref.~\cite{FS1} the spectrum of this radiation is found from the theory of thermal fluctuations in the degenerate positron gas. For a wide range of temperatures of antiQNs, this spectrum is similar to the blackbody radiation spectrum with the frequency-dependent suppression coefficient of order $6\times 10^{-4}$. The temperature of this radiation may be found by imposing the condition of energy balance between the radiated and absorbed energy due to interaction with the ambient medium. It is important to notice that this interaction is drastically different for QNs and antiQNs. Indeed, atoms and molecules are supposed to scatter mostly elastically off QNs while they are annihilated upon collision with an antiQN. The latter assumption may be too strong, and, more generally, we introduce a phenomenological coefficient $\kappa$, $0<\kappa<1$, which measures the probability of annihilation of atoms or molecules upon collision with antiQN, so that $\kappa=0$ corresponds to ordinary QNs which interact with visible matter elastically. 

Although quark matter at zero or low temperature has been hypothesized long time ago as a possible state of matter deep inside neutron stars \cite{QuarkStar}, it has never been observed directly, and many its properties remain unclear. Yet less is known about antiquark matter and its interactions with baryonic matter. One feasible option is that nucleons annihilate near the surface of antiQN. This assumption was considered in the works \cite{FS0,FS1,FS2}.  Indeed, the attenuation length due to elastic scattering of a nucleon in nuclear matter at the relevant energy is equal to $0.3$ fm \cite{FS0}, so this seems natural to assume that atomic nuclei cannot penetrate deep inside the core and annihilate on the boundary of antiquark core with emission of charged and neutral pions. 

There is a nonzero probability of an incident nucleus reflecting off the antiQN surface and escaping without annihilation. However, the nucleus loses part of its kinetic energy due to friction in the positron cloud, making it unlikely to overcome the Coulomb barrier of the antiQN. Additionally, some energy may be transferred into excitations of the nucleus and the antiQN. A feasible scenario in this case is that the nucleus bounces near the antiQN surface until the annihilation process starts \cite{FS0}. The energy released from the annihilation of the first nucleons destroys the nucleus and may provide escape energy for the remaining nucleons that move away.

A possible escape mechanism for hydrogen nucleus was suggested in Ref. \cite{Forbes2010}: transformation of proton to neutron in the charge exchange reaction with antiQN. In the case of  a heavier nucleus, reduction of its charge may also work.  However,  in the conventional nuclear physics at low energies the annihilation process is dominant over the charge exchange and large angle elastic scattering, see, e.g., Ref. \cite{Richard}.

An opposite conceivable scenario suggests low efficiency of annihilation of antiquark matter with the baryonic matter, so that nucleons can penetrate deep inside the antiquark core due to possible superfluidity of (anti)quark matter in the color-superconducting phase and eventually annihilate with no emission of pions outside of the quark core because the decay products may be absorbed via the strong interaction. In the latter case, $2m_p c^2$ of energy per each matter nucleon is transferred mainly into the heat of the positron cloud. 

To take into account both these cases, we introduce a parameter $\xi$, $0\leq \xi\leq1$ which measures the fraction of the incident mass which is transferred into free pions outside the antiquark core. Weak decay of charged pions produces neutrinos which may be detected.

The authors of Ref.~\cite{Lawson2015} propose the third scenario when nucleon annihilation produces pionlike (or kaonlike) low mass quasiparticles, $m_{\pi} \approx 5 - 20$ MeV, deep inside antiquark core. It was proposed that weak decay of such pions may only produce low-energy neutrinos with energies below the detection threshold of neutrino detectors (or below the conventional Sun neutrinos background). However, the weak decay time of such low-mass pions should be many orders of magnitude bigger than the lifetime due to the strong interaction with the antiquark media (i.e., due to pion absorption), with further transfer of the energy to positron cloud by electromagnetic interaction. Indeed, it is not enough energy to produce muons while decay of such pion with production of electron is very strongly suppressed. Thus, we believe that this mechanism may produce only a negligible fraction of neutrinos.

To summarize, $A$, $\kappa$ and $\xi$ are three basic unknown parameters in the antiQN model which will be studied in this paper. In most of the formulas we will use natural units with $\hbar=c=1$, but represent the results of out estimates in physical units.


\section{Manifestation of antiquark nugget  events in dense media}
We begin this section with a brief review of the spectrum of thermal radiation from antiQNs at high temperatures, following Ref.~\cite{FS1}. We show that antiQNs emit $\gamma$-rays when they interact with dense media such as the Earth's crust and core. Since the absorption length of $\gamma$-rays is very short in dense media, we argue that antiQNs are capable of producing explosionlike events when they move through the Earth. We then reconsider the results of seismic detection of DM particles to find the limit on the parameter space in the antiQN model of dark matter.

\subsection{Thermal radiation from antiquark nugget in a rock}

When antiQN propagates through a dense medium, matter particles annihilate inside the antiquark core and contribute to the temperature $T$ of the positron cloud. The power of thermal radiation per unit angular frequency interval $d\omega$ per unit surface area of a spherical particle at temperature $T$ may be represented as (see, e.g., Ref.~\cite{BohrenHuffman})
\begin{equation}
    P(\omega,T) =  \pi I_0(\omega,T) E(\omega)\,,
    \label{P}
\end{equation}
where $I_0(\omega,T) = \frac1{4\pi^3}\frac{\omega^3}{\exp(\omega/T)-1}$ is the Planck function and $E(\omega)$ is the thermal emissivity of the spherical particle. The latter was studied in Ref.~\cite{FS1} for the (anti)quark nuggets. In particular, for large frequency $\omega\gg c/R_\text{QN}$, the thermal emissivity of antiQN may be approximated by its asymptotic expression \cite{FS1}
\begin{equation}
    E(\omega) \approx 5.36\,\text{Re}[\zeta(\omega)]\,,
\end{equation}
where $\zeta(\omega) = \sqrt{\varepsilon_0/\varepsilon(\omega)}$ is the impedance, $\varepsilon_0=1$ is the vacuum permittivity and
\begin{equation}
    \varepsilon(\omega) = 1 -\frac{\omega_p^2}{\omega^2 + i\gamma \omega}
\end{equation}
is the relative permittivity (dielectric constant). Here $\omega_p\approx 2$\,MeV is the plasma frequency and $\gamma \approx475$\,eV  is the damping constant of the degenerate positron gas adjacent to the antiQN core. 

Integration of the function (\ref{P}) over the frequency yields the total radiation power per unit surface area of antiQN, $W(T) = \int_0^\infty P(\omega,T)d\omega$. This power may be conveniently represented as
\begin{equation}
    W(T) = R(T) \sigma_\text{SB} T^4\,,
    \label{W}
\end{equation}
where $\sigma_\text{SB} = \frac{\pi^2}{60}$ is the Stefan–Boltzmann constant (in natural units) and $R(T)<1$ is the function describing deviation of the radiation power from the blackbody radiation. In Ref.~\cite{FS1} it was shown that this function varies from $10^{-4}$ to $10^{-3}$ for a wide range of temperatures from $T=1$\,eV to $T=100$\,keV.

An antiQN moving through a medium gains the energy mainly due to annihilation of nucleons inside the antiquark core. The energy release per each annihilated nucleon is about $2m_p$, with $m_p$ for the nucleon mass. Thus, the incoming energy flux due to the nucleon annihilation is 
$W_\text{in} = 2 m_p \sigma_\text{ann}v n_\text{matter}=2 \rho_\text{matter}\sigma_\text{ann}v$,
where $\sigma_\text{ann}\approx \kappa \pi R_\text{QN}^2$ is the annihilation cross section,\footnote{As is demonstrated in Appendix \ref{AppB}, photon pressure on the atoms of matter plays minor role for antiQNs with $A<10^{35}$. Therefore, we ignore the contribution from this effect to the annihilation cross section.} $v$ is the antiQN speed relative to the medium and $n_\text{matter}$ nucleon number density in the medium. The speed $v$ should be taken from a distribution of velocities of DM particles in the Galaxy, but for rough estimates we take $v\approx 250$\,km/s the mean DM particle velocity in the Solar neighborhood. 

The radiated energy flux from antiQN is proportional to its surface area, $W_\text{rad} = 4\pi R_\text{QN}^2 W(T)$ with $W(T)$ given in Eq.~(\ref{W}). When an antiQN propagates through a homogeneous medium, it quickly acquires the thermal equilibrium with temperature $T$ specified by the energy balance, $W_\text{in}=W_\text{rad}$, or
\begin{equation}\label{eqT}
    \kappa (1-\xi) \rho_\text{matter} v=2 R(T) \sigma T^4  
\end{equation}
This equation was solved numerically in Ref.~\cite{FS1} for different media. In particular, for $\kappa (1-\xi)=1$ antiQN temperature in a rock is found in the interval between 80\,keV and 90\,keV depending on the density of rock. For specific estimates we will further take the value 
\begin{equation}
     T\approx [\kappa (1-\xi)]^{1/4} 87\text{ keV},
    \label{T}
\end{equation}
although more generally a range of temperatures may be considered. Note that temperature $T$ has a very weak dependence on the parameters $\kappa$ and  $(1-\xi)$. The total emitted energy is actually determined by the equilibrium condition (\ref{eqT}). In this paper, temperature $T$ is needed only to determine the spectrum of thermal radiation from antiQNs.

\begin{figure}
    \centering
    \includegraphics[width=8cm]{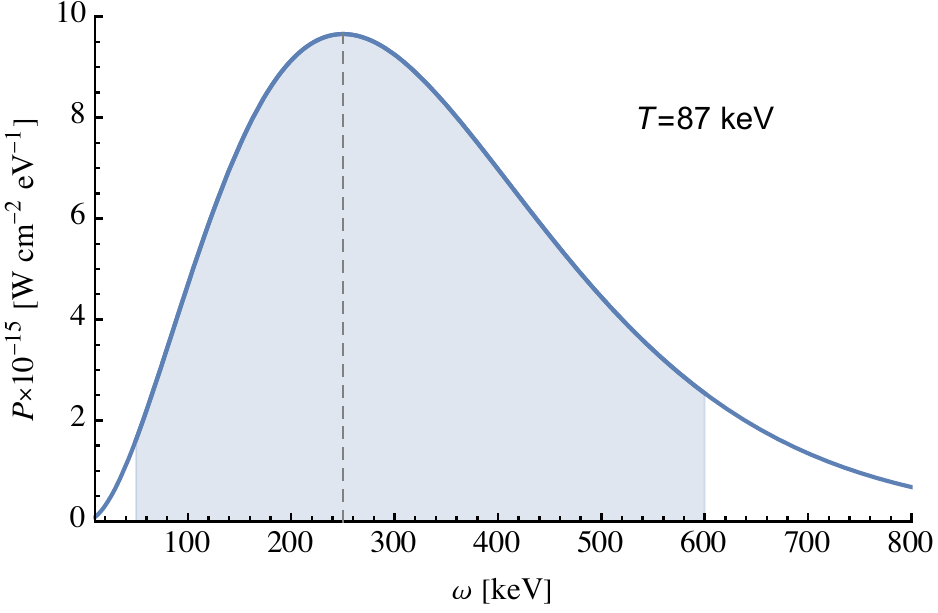}
    \caption{Spectrum of thermal radiation from antiQN at temperature $T=87$\,keV. Shaded area in the region from 50 to 600 keV covers about 90\% of thermal radiation power. Dashed vertical line at $\omega=250$\,keV indicated the maximum of thermal radiation power.}
    \label{fig:Power}
\end{figure}

It is instructive to consider the spectrum of thermal radiation from antiQN at the temperature (\ref{T}), as described by the function (\ref{P}), see Fig.~\ref{fig:Power}. At $T=87$\,keV 90\% of radiation power are confined in the interval between 50 and 600 keV, and the spectrum peaks at $\omega=250$\,keV. Thus, most of the radiation from antiQN in a dense medium like a rock corresponds to gamma-ray spectrum. Typical attenuation length of such photons in a rock is about 1-5 cm, see, e.g., Ref.~\cite{AttenuationLength} for gamma-ray absorption in SiO$_2$. Thus, over 90\% of thermal radiation from antiQN moving through a rock is absorbed in a cylinder with radius 5 cm.

\subsection{Comparison of antiQN collision events with explosions}
\label{SecExplosives}

Let us estimate the amount of energy emitted by antiQN per $l=1$\,m of its trajectory in a rock. This energy is simply given by the annihilation of matter in the volume swept by the antiQN, $E=2\rho_\text{rock}\sigma_\text{ann}l\approx 2\pi \kappa l(1\,\text{fm})^2 \rho_\text{rock} A^{2/3}$, where the annihilation cross section is proportional to the  geometric cross section, $\sigma_\text{ann}\approx \pi\kappa (1\,\text{fm})^2 A^{2/3}$. Indeed, atoms swept by the antiQN trajectory are likely to collide with antiQN because the antiQN's speed exceeds the sound and thermal velocity of atoms by two orders of magnitude. Thus, antiQN collision events with dense medium like a rock should release huge amount of energy for $A>10^{25}$, see Table \ref{tab:TNT}. In particular, for $A=10^{26}$ and $\kappa=0.1$, the total energy radiated by antiQN per 1\,m of its trajectory in a rock is about 0.032 GJ which is equivalent to the energy release from about 7.7 kg of TNT explosives. We stress that this comparison is relevant because the radiation from antiQN in a rock is absorbed in a small volume around the antiQN trajectory thus creating an overpressure comparable to the real explosives. Although this comparison is just qualitative, it demonstrates that interaction of antiQN with any dense medium like a rock or soil should produce very powerful explosionlike events which cannot remain unobserved if they happen on a populated area.

\begin{table}[tb]
\centering
    \begin{tabular}{c|c|c|c}
 $A$ & $E$/TJ & TNT equivalent, ton & $\langle \dot N\rangle$/(100yr)/(0.01 $S_\oplus$) \\\hline
 $10^{25}$ & $6.9\times 10^{-3}$ & 1.7 & $2\times 10^7$ \\
 $10^{26}$ & 0.032 & 7.7 & $2\times 10^6$ \\
 $10^{27}$ & 0.15 & 36 & $2\times 10^5$\\
 $10^{28}$ & 0.7 & 170 & $2\times 10^4$\\
 $10^{29}$ & 3.2 & 770 & $2\times 10^3$\\
 $10^{30}$ & 15 & 3600 & 200\\
 $10^{31}$ & 70 & $17\times 10^3$ & 20\\
 $10^{32}$ & 320 & $77\times 10^{3}$ &2\\
 $10^{33}$ & 1500 & $36\times 10^{4}$ &0.2
    \end{tabular}
\caption{Energy release $E$ per one kilometer of antiQN trajectory in a rock with density $\rho_\text{rock}=2.7$\,g/cm$^3$ for different baryon charge numbers from $A=10^{25}$ to $A=10^{33}$. Here we assume a conservative estimate $\kappa=0.1$ for the coefficient describing efficiency of annihilation. The TNT energy release equivalent is 4184 J/g. $\langle \dot N\rangle$ is mean frequency of antiQN events per 1\% of Earth's surface area per a 100-year period.}
    \label{tab:TNT}
\end{table}

The average collision frequency of antiQNs with the Earth was estimated in Ref.~\cite{Lawson1,Budker2020} assuming that they saturate the local dark matter density $\rho_\text{DM} = 0.3\,\text{GeV}/\text{cm}^3$,
\begin{equation}
    \langle \dot N\rangle \simeq 2.1\times 10^7 \text{year}^{-1} \left(\frac{10^{25}}{\langle A\rangle} \right).
    \label{Ndot}
\end{equation}
Note that the populated area of the Earth comprises about 0.9\% of total surface area of the Earth. It is reasonable to assume that unexplained explosions on the populated area of the Earth have not been observed for at least 100 years (keeping in mind the mysterious Tunguska event in 1908). Requiring that $\langle \dot N\rangle <1$ on this timescale can  become a strong constraint on the baryon charge $A$,
\begin{equation}
    A> 5\times 10^{32}\,.
    \label{Blimit1}
\end{equation}
Note that in this constraint we have not assumed any mechanisms potentially suppressing annihilation of visible matter with antiQNs except for $\kappa\sim 0.1$. Indeed, any of such mechanisms requires additional assumptions about the physics of antiQNs which may be debated. Note that for $A> 10^{30}$ the energy released on the last 100 meters under Earth surface may exceed that of atomic bomb.

If the efficiency of annihilation coefficient is small enough, $\kappa \sim  0.1$, the values of $A<10^{26}$ cannot be excluded by this qualitative consideration (interval $0.1< \kappa<0.5 $ is a natural range for the annihilation of a matter nucleus near antiQN surface since annihilation of the first nucleons causes explosion of the nucleus with probable escape  of at least half of remaining nucleons which move outside).  For an arbitrary value of $\kappa$ we may  present  a rough estimate for the upper bound on the baryon charge number as:
\begin{equation}
    A\kappa <10^{25}\,.
\label{fountains-limit}
\end{equation} 
Note that for higher value of $A$,  observable fountain of hot gases may come from the depth far exceeding 1 m and may significantly increase visible effect of explosion. In addition, antiQN stays hot and continues to radiate x-rays few meters above the ground until its temperature comes to new equilibrium value $T \sim 1$ keV in atmosphere. 

In Ref.~\cite{Budker2020} passage of antiQN was linked to skyquakes, loud explosionlike sounds of unknown origin reported by many people in different regions around the world. However, no observations of fire fountains and holes in the ground associated with  skyquakes have been reported.

\subsection{Limits from seismic data}
\label{SecSeismic}

The above comparison of antiQN events on Earth with explosives is rather qualitative, and the limit (\ref{Blimit1}) needs to be further justified by stronger quantitative results. Such results can be found from reanalysis of the works \cite{ApolloConstraint,Seismic1,Seismic2} where signals of transitions of quark nuggets through the Earth and Moon were studied. Unlike the QN model, the antiQN one assumes that the main channel of energy deposition from DM particle collision with visible matter is through the matter-antimatter annihilation, $\Delta E_\text{ann} = 2\kappa m_\text{mat}c^2$, for a matter particle (atom or molecule) of mass $m_\text{mat}$.  In contrast, the maximum amount of energy transfer in elastic collision is   $\Delta E_\text{coll} \approx 2m_\text{mat}v_\text{DM}^2$,
in the approximation $m_\text{mat}\ll m_\text{DM}$. Thus, antiQNs deposit at least  $\kappa 10^6$ times more energy than QNs,
\begin{equation}
    \frac{\Delta E_\text{ann}}{\Delta E_\text{coll}}\approx \kappa\frac{c^2}{v^2}\approx \kappa 10^6\,.
    \label{EnergyRatio}
\end{equation}

The authors of the works \cite{Seismic1,Seismic2} studied the possibility of detection of QNs with seismic detectors on the Earth. The distinguishing feature of such events should be a specific wavefront at different locations on the Earth created by linear supersonic motion of massive DM particles through the Earth. This wavefront should create specific signal in a distributed network of seismic detectors on the Earth. Nonobservation of this signal allowed the authors of Refs.~\cite{Seismic1,Seismic2} to practically rule out the QNs with masses over 10 metric tons. Indeed, assuming that about 1\% of energy is transformed into seismic waves of frequency about 1 Hz, it was shown that the seismic detectors are capable to detect DM particles with mass 
\begin{equation}
    m_\text{DM}\gtrsim 13\times 10^{6}\, \text{g}\approx 7\times 10^{30}\,\text{GeV}.
\end{equation}
Making use of the relation (\ref{EnergyRatio}) we conclude that the same set of detectors is sensitive to antiQNs with mass $m_\text{QN}>13\kappa^{-1}$\,g, or, assuming that $m_\text{QN}\simeq A m_p$, we get the upper bound on the baryon charge number:
\begin{equation}
    A\kappa <8\times10^{24}\,.
\label{earthquakes-limit}
\end{equation}

The results of the works \cite{Seismic1,Seismic2} were based on the data of seismic activity on the Earth in the period of about four years from 1990 to the end of 1993. This implies that antiQN collision frequency with the Earth should be less than one event per four years, $\dot N< (4\text{ years})^{-1}$. Making use of Eq.~(\ref{Ndot}) this translates into the following constraint on $A$:
\begin{equation}
    A > 8\times 10^{32}\,.
\end{equation}
This appears even stronger than the rough estimate (\ref{Blimit1}).

The constraint (\ref{earthquakes-limit}) strongly depends on the assumption made in Refs.~\cite{Seismic1,Seismic2} that the quark nugget can transfer about 1\% of its kinetic energy to seismic waves with frequency about 1 Hz and about 5\% of its kinetic energy to seismic waves with frequencies under 20 Hz. This assumption was debated in Ref.~\cite{Starkman9} where it was shown that the energy of seismic waves on the Moon from quark nuggets was strongly overestimated. 

More precisely, the estimated fraction of kinetic energy of a QN with cross section $\sigma$ that may be transferred to the seismic waves with frequencies under 20 Hz is  $\Xi_\mathrm{QN} < 4.5\times(\sigma/\mathrm{cm}^2)$ \cite{Starkman9}. In the case of antiQN, however, this quantity is enhanced by the factor (\ref{EnergyRatio}) because the energy is released upon annihilation of matter atoms and molecules colliding with antiQN. For the annihilation cross section $\sigma_\mathrm{ann} = \pi\kappa \mathrm{fm}^2 A^{2/3}$ we find
\begin{equation}
    \Xi_\mathrm{antiQN} < 2\times 10^{-19} \kappa A^{2/3}\,.
    \label{XiantiQN}
\end{equation}
In particular, for $\kappa=1$ and $A=10^{26}$ the quantity in the right-hand side reaches the value 0.04. Thus, it is not excluded that up to 4\% of annihilation energy may be converted into seismic waves upon collisions of antiQN with Earth's surface. Equation (\ref{XiantiQN}) may, however, overestimate the energy of seismic waves because the authors of the work \cite{Starkman9} gave an upper bound on this quantity based on the results of the Apollo mission on the Moon. An accurate estimate of this quantity for antiQNs hitting the Earth requires a separate study which will be given elsewhere.


\section{Neutrino flux from antiQN annihilation}

As was shown in Ref.~\cite{Gorham_2017}, large celestial bodies can produce a significant flux of neutrinos due to matter annihilation in the antiQN DM particles. The leading contributions to this flux are sourced by the Sun and by the Earth. Although neutrinos of all flavors are produced in this process, we will focus on the electronic antineutrino because the Super-Kamiokande detector has the highest sensitivity to them after subtraction of the ordinary solar neutrino background. The experimental upper limit on the electronic antineutrino flux with energy $E>17.3$~MeV is \cite{DSNB_SuperK_2021}
\begin{equation}
    \Phi_{\bar\nu_e,\rm exp} = 2.7 \text{ cm}^{-2}\text{s}^{-1}\,.
    \label{exp-limit}
\end{equation}
In this section, we will estimate the contribution to this flux from annihilation of antiQNs in the Sun and in the Earth and compare it with the experimental upper limit (\ref{exp-limit}).

\subsection{Neutrinos from the Sun}

We consider annihilation of matter in collisions with antiQNs following Refs.~\cite{FS0,FS1,FS2}, i.e., similar to the nucleon-antinucleon annihilation where pions give primary decay products. On average each proton-antiproton annihilation event produces two neutral pions and three charged ones. In a dense medium like the interior of the Sun $\pi^-$ mesons are captured and then absorbed by nuclei and annihilate through strong interaction with no neutrino emission. $\pi^+$ mesons decay through weak interaction with production of the following leptons: $\pi^+\to \mu^+\nu_\nu \to e^+\nu_e\bar\nu_\mu\nu_\mu$. Thus, each proton annihilation in the antiQN produces about 1.5 muonic antineutrino if all pions escape from the antiquark core. 

However, if pions are produced deep inside the antiquark core, only a fraction of these pions can leave the antiquark core with subsequent decay into leptons. This effect is taken into account by the yet undetermined coefficient $\xi<1$. Thus each matter proton annihilating in the antiQN can produce about $1.5\xi$ muonic antineutrinos. The pions which are confined inside the quark core annihilate eventually via strong interaction with no neutrino emission. The coefficient $(1-\xi)$ may be considered as the thermalization coefficient which measures the thermalized fraction of total $2m_pc^2$ energy.

To estimate the production rate of $\bar\nu_\mu$ in the Sun we first note that the capture cross section is enhanced by the gravitational attraction,
\begin{equation}
\label{capture}
    \sigma_\text{cap} = \pi R_\odot^2 \left(1 + \frac{2G M_\odot}{R_\odot v^2} \right),
\end{equation}
where $M_\odot$ and $R_\odot$ are mass and radius of the Sun, respectively, $G$ is the gravitational constant, and $v\sim250$\,km/s is the DM particle velocity at infinity. Once captured in the Sun, antiQNs should fully annihilate with emission of about $1.5\xi A$ muonic antineutrinos among other decay products. Assuming that the fraction of antiQNs in all dark matter is about $3/5$, we estimate the production rate of $\bar\nu_\mu$:
\begin{equation}
\label{production}
    \Gamma_{\bar\nu_\mu} \simeq 1.5\xi A v \frac35\frac{\rho_\text{DM}}{m_\text{QN}} \sigma_\text{cap}\,,
\end{equation}
where $\rho_\text{DM}\simeq 0.3$\,GeV/cm$^3$ is the local dark matter density and $m_\text{QN}\simeq A m_p$ in the mass of one antiQN.

The muonic antineutrinos produced in the Sun may be detected on the Earth as $\bar\nu_e$ because of the neutrino oscillation. The probability of this oscillation in vacuum is $P_\text{vac}(\bar\nu_\mu\to\bar\nu_e)\approx 2/9$, see, e.g., Ref.~\cite{WIMPFlavour}. In a dense medium like the Sun's core this probability is slightly different due to the Mikheyev-Smirnov-Wolfenstein effect, $P_\text{mat}(\bar\nu_\mu\to\bar\nu_e)\approx 1/6$. This contributes to the theoretical uncertainty of our estimates, but for a conservative estimate we accept the value $P=P_\text{mat}(\bar\nu_\mu\to\bar\nu_e)\approx 1/6$. Taking into account this probability we find the electron antineutrino flux on Earth:
\begin{equation}
    \Phi_{\bar\nu_e} = \frac{\Gamma_{\bar\nu_\mu} P}{4\pi L^2}\,,
\end{equation}
where $L=1$\,a.u.\ is the mean distance between the Sun and the Earth. Making use of Eqs.~(\ref{capture}) and (\ref{production}), we find that the electronic antineutrino flux on the Earth is independent of the antiQN mass number $A$ in the leading-order approximation if $A\leq 10^{30}$,
\begin{align}
    \Phi_{\bar\nu_e} &\simeq \frac{9}{40}\frac{\xi v \rho_\text{DM} R_\odot^2}{m_p L^2}\left(1 + \frac{2G M_\odot}{R_\odot v^2} \right) P\nonumber\\
     &\approx 46 \xi \,\mbox{cm}^{-2}\mbox{s}^{-1}\,.
\label{flux-Sun}
\end{align}
This value is compatible with the experimental limit (\ref{exp-limit}) for $\xi< 0.06$. 

If $A>10^{30}$ some of antiQNs may traverse the Sun without complete annihilation and escape. In Appendix \ref{appA} we consider trajectories of antiQNs with mass number $10^{31}<A<10^{37}$ incident on the Sun with variable impact parameter $b\in[0,R_\odot]$ in the vicinity of the Sun. For each value of $A$ we define a critical value of the impact parameter $b_\mathrm{crit}(A)$ so that antiQNs impacting the Sun with $b>b_\mathrm{crit}$ are likely to traverse the Sun and exit with speed higher than the escape velocity near the Sun's survace, $v>v_\mathrm{esc}=617$\,km/s. Such antiQNs lose a small fraction of their mass and do not give significant contributions to the neutrino flux on the Earth. Thus, for $A>10^{30}$ the neutrino flux (\ref{flux-Sun}) is reduced by the factor $\varsigma \equiv (b_\mathrm{crit}/R_\odot)^2$. In particular, for $A=10^{33}$ we estimate $\varsigma \approx 0.35$, while for other values of $A$ this parameter may be described by the following fitting function
\begin{equation}
\varsigma=\left\{
\begin{array}{l}
1  \quad\mbox{ for } A\leq 10^{30}\\
3.15-0.034\ln A  \quad\mbox{ for } 10^{30}< A\leq 10^{37}\,.
\end{array}
\right.
\end{equation}
We conclude that the observed neutrino flux (\ref{exp-limit}) imposes the following limit on the parameter $\xi$:
\begin{equation}
    \xi< 0.06 \varsigma^{-1}\,.
\end{equation}


\subsection{Neutrinos from the Earth}
\label{SecEarth}

The calculation of the neutrino flux from annihilation of antiQNs in the Earth is similar to the one in the Sun, but with a few important differences. 

(i) The gravitational attraction of the Earth is much weaker than that of the Sun. Thus, the antiQN collision cross section with the Earth is very close to the Earth's geometric cross section $\sigma_\oplus = \pi R_\oplus^2$. 

(ii) AntiQN particles have a relatively small cross section of interaction with matter, and a high speed relative to the Earth, $v\sim 250$\,km/s. It is easy to check that for antiQNs with $A\gtrsim10^{24}$ the stopping power due to collisions with the Earth is small, and trajectories of DM particles remain to be close to straight lines, see Appendix \ref{appA}. As a result, each antiQN can lose only a small fraction of its mass due to interaction with the matter particles of the Earth, in contrast with the antiQN annihilation in the Sun, where they are likely to fully annihilate. 

(iii) The matter density of the Earth is much smaller than that of the Sun's core. As a result, for muonic to electronic antineutrino oscillations with energy under 30 MeV the Mikheyev-Smirnov-Wolfenstein effect plays minor role and may be ignored. Thus, for our estimates we can simply take the vacuum neutrino oscillation probability, $P=P_\text{vac}(\bar\nu_\mu\to\bar\nu_e)\approx 2/9$.

Considering the above simplifications, the mean number density of antiQNs inside the volume of the Earth at each instance is
\begin{equation}
\label{number-density}
    n_\text{QN} \approx  \frac{3}{5}\frac{\rho_\text{DM}}{m_\text{QN}}\,,
\end{equation}
where the factor 3/5 takes into account the fraction of antiQNs in all dark matter. Each of these antiQN particles collides with atoms and molecules in the Earth and annihilates them. Remembering that the annihilation cross section for matter on antiQN is suppressed by the yet undetermined coefficient $\kappa$, $\sigma_\text{ann} = \kappa \pi R_\text{QN}^2$, the annihilation rate may be estimated as
\begin{equation}
\label{ann-rate}
    \sigma_\text{ann} v \frac{\rho_\oplus}{m_p}\,,
\end{equation}
where $\rho_\oplus$ is the position-dependent density of the Earth. 

Assuming that about $1.5\xi$ muonic antineutrinos are produced per each nucleon annihilation in the antiQN, we integrate the expression (\ref{ann-rate}) with the antiQN number density (\ref{number-density}) over the volume of the Earth $V_\oplus$ to estimate the total muonic antineutrino production rate,
\begin{align}
    \Gamma_{\bar\nu_\mu} &\simeq 1.5\xi \int_{V_\oplus} \sigma_\text{ann} v \frac{\rho_\oplus}{m_p} \frac{3}{5}\frac{\rho_\text{DM}}{m_\text{QN}} dV
    \nonumber\\
    &=0.9\xi\kappa v \pi R_\text{QN}^2\frac{\rho_\text{DM}M_\oplus}{m_p m_\text{QN}}\,,
    \label{muonic-rate-Earth}
\end{align}
where $M_\oplus$ is the mass of the Earth. Finally, the electronic antineutrino flux at a detector on the surface of the Earth is obtained by multiplying Eq.~(\ref{muonic-rate-Earth}) by the average neutrino oscillation probability $P$ and by dividing it by the Earth' surface area,
\begin{equation}
    \Phi_{\bar\nu_e} = \frac{\Gamma_{\bar\nu_\mu}P}{4\pi R_\oplus^2} 
    =0.9\xi\kappa v P \frac{R_\text{QN}^2 M_\oplus \rho_\text{DM}}{4R_\oplus^2 m_p m_\text{QN}}\,.
\end{equation}

To give a numerical estimate of the electronic antineutrino flux on the Earth, we recall that $R_\text{QN}\simeq 1\,\text{fm}A^{1/3}$ and $m_\text{QN}=A m_p$. As a result, we find
\begin{align}
    \Phi_{\bar\nu_e} &= 0.9 \xi\kappa A^{-1/3} v (1\,\text{fm})^2 P\frac{M_\oplus \rho_\text{DM}}{4 R_\oplus^2 m_p^2}\nonumber\\
    &\approx \xi\kappa A^{-1/3} 3.5\times 10^{13}\,\mbox{cm}^{-2}\mbox{s}^{-1}\,.
\label{flux-Earth}
\end{align}
Note that in contrast with the corresponding flux from the Sun (\ref{flux-Sun}), Eq.~(\ref{flux-Earth}) has nontrivial dependence on the antiQN mass number $A$.

The estimated neutrino flux (\ref{flux-Earth}) does not exceed the experimental bound (\ref{exp-limit}) if the parameters of the model obey
$\xi\kappa A^{-1/3} < 7.7\times 10^{-14}$.


\section{Discussion}

\subsection{Analysis of constraints}
\label{SecAnalysis}

The main contributions to the electronic antineutrino flux are given by Eqs.~(\ref{flux-Sun}) and (\ref{flux-Earth}). In general, we have to combine these two contributions and compare them with the experimental bound (\ref{exp-limit}). For a conservative estimate, we can require that each of these contributions does not exceed the bound (\ref{exp-limit}). Remembering also the limit from nonobservation of antiQN-induced earthquakes (\ref{earthquakes-limit}) we arrive at the following set of the limits on the parameters of the model
\begin{subequations}
\label{allConditions}
\begin{align}
    A \kappa &<8\times 10^{24} \label{conditionA}\\
    \xi\kappa A^{-1/3} &< 7.7\times 10^{-14} \label{conditionB}\\
     \xi&< 0.06 \varsigma^{-1}\,.  \label{conditionC}
\end{align}
These constraints should be augmented with the condition of survival of antiQNs in the hot early Universe from Ref.~\cite{FS2}
\begin{equation}
\kappa A^{-1/3} < 0.8 \times 10^{-9}\,.
\label{conditionD}
\end{equation}
\end{subequations}

Mathematically, all conditions (\ref{allConditions}) are consistent. However, for physical reasons, the parameter $\kappa$ cannot be arbitrarily small because it effectively measures the nucleon annihilation probability upon collision with an antiQN. For comparison, the antiproton annihilation cross section in collisions with heavy nuclei is suppressed by a factor of $\kappa \approx 0.6$ compared to the total scattering cross section (see, e.g., Refs.~\cite{Richard,antiproton2}). Although the antiquark core is supposed to be in a color-superconducting state, which should have a lower ground state energy compared to standard nuclear matter, the coefficient $\kappa$ is unlikely to be smaller than 0.1,
\begin{equation}
    \kappa\gtrsim 0.1\,,
    \label{kappaPhisical}
\end{equation}
because of the  strong Coulomb attraction of the proton to the antiquark core. 
According to estimates in Ref.~\cite{FS1}, an accelerated proton loses a significant fraction of its kinetic energy due to collisional friction within the positron cloud and becomes bound. Therefore, it cannot escape and must either annihilate or turn into a neutron upon collision with the antiquark core. More accurate estimates of the parameter $\kappa$ require a many-body simulation of this process in the specific model of the  antiquark core which goes beyond the scope of this paper. Note that in Ref.~\cite{Budker2020} it was argued that the value $\kappa\sim1$ is natural in the case of interaction of antiQNs with baryonic matter in a neutral atomic or molecular state while this parameter can develop very small values, $\kappa<10^{-3}$, in the early Universe in the period of big bang nucleosynthesis studied in Ref.~\cite{Survival}.

Keeping in mind the additional constraint (\ref{kappaPhisical}), we can systematically analyse the conditions (\ref{allConditions}). 

First, we note that the conditions (\ref{conditionA}) and (\ref{conditionD}) are inconsistent for $\kappa>0.25$. Combining this condition with (\ref{kappaPhisical}), we have
\begin{equation}
    0.1\lesssim \kappa\leq 0.25\,.
    \label{kappaWindow}
\end{equation}
For each value of $\kappa$ in this interval, the antiQN mass number is constrained by
\begin{equation}
    2\times 10^{27} \kappa^3 < A < 8\times 10^{24} \kappa^{-1}\,.
    \label{Awindow}
\end{equation}

For instance, if $\kappa=0.1$, the mass number may vary within the relatively large window, $2\times 10^{24}<A<8\times 10^{25}$. As is argued above, it is likely, however, that $\kappa$ is not small, and may rich $\kappa=0.25$. In this case, the mass number has a fixed allowed value, $A = 3.2\times 10^{25}$. Note that this is a mean mass number, and the actual antiQNs should have a distribution around this mean value. Finally, we note that if further studies of the quark matter reveal that the nucleon annihilation probability is $\kappa>0.25$, the constraints (\ref{Awindow}) become incompatible, and the antiQN model of dark matter may be excluded for any $A\lesssim 10^{33}$.

Once the region of allowed values in the space $(\kappa,A)$ is fixed, the values of the parameter $\xi$ are specified by the conditions (\ref{conditionB}) and (\ref{conditionC}). The strongest limit on $\xi$ appears at $\kappa=0.1$ and $A=2\times 10^{24}$: $\xi<9.7\times 10^{-5}$. The most conservative limit is for $\kappa=0.1$ and $A=8\times 10^{25}$
\begin{equation}
    \xi < 3.3 \times 10^{-4}\,.
    \label{xiWindow}
\end{equation}
For these values of $\xi$ the constraint (\ref{conditionC}) is satisfied.

The limit (\ref{xiWindow}) shows that neutrino emission from matter annihilation on antiQNs should be strongly suppressed. This may be the case if nucleons penetrate deep inside the antiquark core and annihilate there via the strong interaction with no pion emission. This scenario may be realized if the elastic scattering of the nucleon in the quark core is strongly suppressed, and the quark matter behaves like a superfluid. To justify this scenario an accurate model of the quark matter is needed with precise understanding of the pattern of nucleon interaction with this matter.

The above limits (\ref{kappaWindow})--(\ref{xiWindow}) are applicable for $A<8\times 10^{32}$ since, as noticed at the end of Sec.~\ref{SecSeismic}, in the opposite case the constraint (\ref{conditionA}) does not hold. For $A>8\times 10^{32}$, the events of collision of antiQNs with the Earth become sufficiently rare, so that the limit (\ref{conditionB}) does not hold as well. Thus, very large values of the mass number in the antiQN model
\begin{equation}
    A>8\times 10^{32}
\end{equation}
remain unexcluded in the present study. However, as shown in Sec.~\ref{SecExplosives}, explosions in such rare cases are so strong that they may hardly remain unnoticed during human history. Moreover, with $A$ approaching $10^{40}$, such events may be catastrophic for the life on the Earth causing rapid massive extinction of a significant fraction of life species.

\subsection{Revision of astrophysical constraints and detection proposals}

In papers \cite{FS0,FS1,FS2}, a series of astrophysical observable effects were predicted in the antiQN model of dark matter. Some of these effects strongly depend on the values of parameters $\kappa$, $\xi$ and $A$ of this model. In Sec.~\ref{SecAnalysis}, we found the allowed values of these parameters from the analysis of seismic events on the Earth and neutrino flux from annihilation in the Sun and Earth. The allowed values of parameters $\kappa$ (\ref{kappaWindow}) and $A$ (\ref{Awindow}) are close to the ones assumed within the works \cite{FS0,FS1,FS2}, but the parameter $\xi$ appears very small, see Eq.~(\ref{xiWindow}). This small value shows that free pions cannot be produced in the process of matter annihilation on antiQNs, and nearly all annihilation energy is thermalized in the antiQN positron cloud. Taking this into account, we revisit some of the astrophysical effects discussed in Refs.~\cite{FS0,FS1,FS2}.

In Ref.~\cite{FS0}, we studied the diffuse radiation from the Galactic center focusing on the 511 keV photon line. This flux may be produced upon electron-positron annihilation in the interstellar medium assuming that antiQN may be a source of positrons. This effect is not forbidden in the case of small values of the parameter $\xi$ as in Eq.~(\ref{xiWindow}). Indeed, the estimated number of these positron is independent of the parameter $\xi$ since they originate from the positron cloud. In Ref.~\cite{FS0} it was found that the flux of the 511-keV photons sourced by the antiQN DM in the Galactic center is compatible with the one observed with the SPI/INTEGRAL detector \cite{INTEGRAL} if $A\lesssim 10^{24}$. This value of the antiQN mass number is consistent with  the limits (\ref{Awindow}) if $\kappa< 0.08$. Thus, while our rough estimate of $\kappa \gtrsim 0.1$ does not rule out the possibility that antiQN may be a source of positrons, this interpretation appears unlikely.

In Refs.~\cite{FS1,FS2}, it was shown that antiQN DM particles in the Galactic bulge or in large molecular clouds may produce a $\gamma$ photon diffuse background which is compatible with the observed flux in the Fermi-LAT telescope \cite{FermiLAT}. These $\gamma$ photons originate from decaying $\pi^0$ mesons produced in the nucleon annihilation on the antiquark core. This process is, however, strongly suppressed by the small coefficient $\xi$ as in Eq.~(\ref{xiWindow}). Therefore, antiQN cannot produce any significant flux of $\gamma$ photons in the Galactic bulge \cite{FS1} or in large molecular clouds \cite{FS2}.

In Refs.~\cite{FS1,FS2,Garry}, we also studied a number of potential manifestations of antiQNs based on the thermal radiation from the positron cloud. All these effects are not impacted by the current study because the thermal radiation is not suppressed, $1-\xi\approx 1$. It is worth mentioning that Ref.~\cite{Garry} suggests using meteor radars for searches of antiQN DM particles due to the fact that the radiation from antiQNs should produce a specific ionized trail in the upper atmosphere. This approach remains relevant and interesting for further searches of DM particles.

The authors of Ref.~\cite{LP} proposed searching for antiQNs with liquid noble gas DM detectors. The signal in these detectors is expected to be formed by showers of particles initiated by $\gamma$-photons from the decays of $\pi^0$ mesons. However, as we show in the present paper, the emission of such pions (and, hence, $\gamma$-photons) should be strongly suppressed by the factor (\ref{xiWindow}). Therefore, the proposal of Ref.~\cite{LP} does not seem efficient for the detection of antiQNs.

\subsection{Comparison with other limits on antiQN DM}

\begin{figure}
    \centering
    \includegraphics[width=1\linewidth]{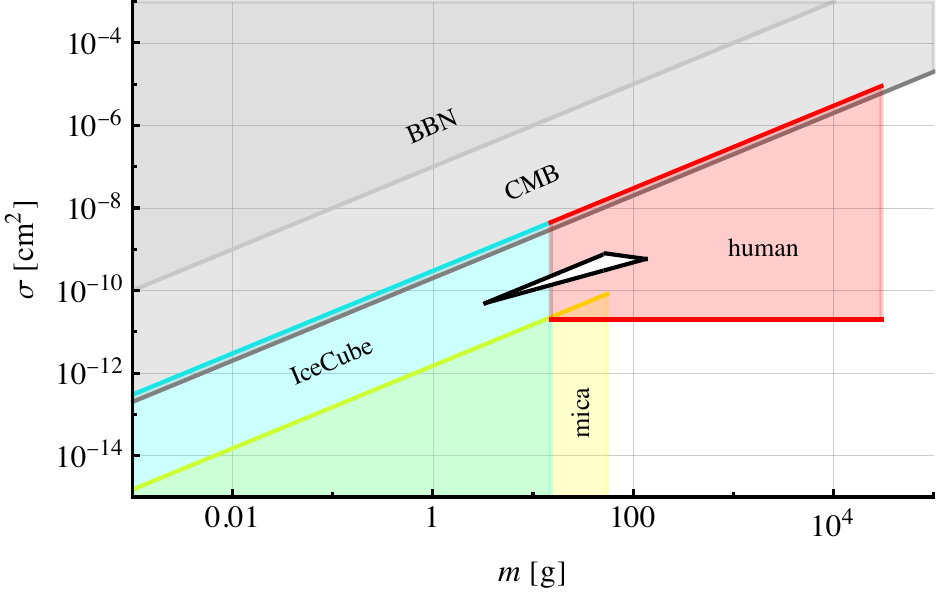}
    \caption{Limits on cross section $\sigma$ of antiQN for masses $m$ in the range from $10^{-3}$ g to $10^5$ g ($A=m/ 1.67 \times  10^{-24}$ g). Gray excluded regions represent limits from CMB and BBN studies in Refs.~\cite{Starkman10,Caloni}. Yellow excluded region comes from a lack of tracks in an ancient slab of mica \cite{Glashow,Price}. Cyan excluded region shows a lack of a signal in the IceCube experiment from Ref.~\cite{Lawson1}. Red excluded area was found in Ref.~\cite{Starkman10} by nonobservations of specific signatures of antiQN impacts in humans. The black triangle represents the nonexcluded region from the present study overlaid on the limits from other works.}
    \label{fig:Exclusion}
\end{figure}

It is instructive to compare our results with the limits on the antiQN dark matter from other works. In Fig.~\ref{fig:Exclusion} we show the nonexcluded window (\ref{kappaWindow},\ref{Awindow}) in the parameter space $(A,\kappa)$ on the plane $(m,\sigma)$ via the relations $m=A m_p$ and $\sigma = \pi\kappa \,\mathrm{fm}^2 A^{2/3}$. This window is represented by the black empty triangle. 

In Fig.~\ref{fig:Exclusion}, we present also the limits on antiQN dark matter from other works including Refs.~\cite{Starkman10,Caloni} with limits from the studies of cosmic microwave background (CMB) and Big Bang nucleosynthesis (BBN); Refs.~\cite{Glashow,Price} with limits from a lack of tracks in an ancient slab of mica; Ref.~\cite{Lawson1} with limits from a lack of a signal in the IceCube experiment and Ref.~\cite{Starkman10} where possible antiQN impacts on humans were considered. Although the latter constraint (shown in red in Fig.~\ref{fig:Exclusion}) covers the nonexcluded area found within this work, we still may leave the possibility of detection of antiQNs with these parameters or complete close off  this parameter space in future studies. 

We stress that our limits on the parameters of the quark nugget model are complementary to the ones found in earlier works \cite{Starkman10,Caloni} because we consider different phenomenological manifestations of antiQN dark matter particles.


\section{Summary and conclusion}

In this paper, we studied the parameter space in the antiquark nugget model of dark matter. We consider three main parameters in this model: (i) $\kappa$, the probability of nucleon annihilation in collision with antiquark core, (ii) $\xi$, the nonthermalized fraction of total annihilation energy which is emitted in the form of pions and their decay products upon collisions of atoms with antiQNs, and (iii) $A$, the mass number of antiQN which appears in the definition of its mass as $m_\text{QN} = A m_p$. We found an allowed window in the space $(\kappa,\xi,A)$ which is consistent with the  following observable effects: (a) absence of specific signatures of events of transitions of antiQNs through the Earth in data of distributed seismic detectors, (b) potentially observable neutrino flux from collisions of antiQNs with the Sun and  Earth, and (c) survival of antiQNs in the hot and dense early Universe considered in Refs.~\cite{FS2,Ge}.

We found that the parameter $\kappa$ is restricted to the narrow interval $0.1\lesssim\kappa<0.25$ whereas $A$ obeys the limits (\ref{Awindow}) which depend on $\kappa$. The largest allowed interval for the mass number $2\times 10^{24}<A<8\times 10^{25}$ corresponds to $\kappa=0.1$. However, these values for $\kappa$ and $A$ may be excluded by other observations, see Fig.~\ref{fig:Exclusion}.

The parameter $\xi$ is limited by the condition (\ref{conditionB}) which depends on $\kappa$ and $A$. The upper limit on $\xi$ for the allowed values of $(\kappa,A)$ appears very strong, $\xi < 3.3 \times 10^{-4}$. It is not clear whether this small value of $\xi$ can be realized in Nature. One possible explanation for this small value could be the effect of the superfluidity of quark matter, which allows nucleons to penetrate deep inside the antiquark core, where it annihilates via the strong interaction without neutrino emission. The energy of annihilation in this case is absorbed by the strong and electromagnetic interactions preventing  weak decays with neutrinos emission. To justify this scenario an accurate model of the  quark matter is needed with precise description of interaction pattern with baryonic matter. This problem is postponed for future studies.

Finally, we note that the obtained in this work limits on the parameter space $(\kappa,\xi,A)$ are valid for $A< 8\times 10^{32}$, see Sec.~\ref{SecSeismic}. In Sec.~\ref{SecExplosives} we have shown that for $A$ beyond this limit the events of collisions of antiQNs with the Earth may be very catastrophic, although rare. For instance, for $A=10^{33}$, the energy release from the interaction of antiQN with a rock along 
100 m of its trajectory may be comparable with 36 kilotons of TNT and hardly would be unnoticed. Therefore, although not fully excluded, such large values of the mass number seem very unlikely to be realized in Nature. Further investigations needed to provide a definitive answer to this question.

\vspace{2mm}
\textit{Acknowledgements.}--- 
The work was supported by the Australian Research Council Grants No.\ DP230101058 and No.\ DP200100150.

\appendix
\section{Trajectories of AntiQNs through Sun and Earth}
\label{appA}

In this Appendix, we study trajectories of antiQNs inside massive bodies like the Sun or the Earth with various initial values of the mass number $A$. Their trajectories are described by classical Newtonian mechanics, though with some peculiarities \cite{Gorham_2017}. 

Let $\mathbf{r}$ be a position vector of antiQN in the reference frame of a massive body (Sun or Earth) with mass $M$, and $\mathbf{v}=\dot{\mathbf{r}}$ be the corresponding velocity vector. Inside the massive body the antiQN experiences the gravitational attraction with the force $-G m_\text{QN}(t) M_\text{int}(r)\mathbf{r}/r^3$, where $M_\text{int}$ is the enclosed mass of the attracting body within the sphere of radius $r$, $M_\text{int} = 4\pi\int_0^r \rho(r)r^2dr$. Here $\rho(r)$ is the density of the attracting body. The gravitational attraction force is partly compensated by the collisional friction given by $-\frac12\sigma_\text{col} \rho(r) v \mathbf{v}$, where $\sigma_\text{col} \simeq \pi R_\text{QN}^2$ is the collisional cross section of antiQN which is supposed to be close to its geometric cross section. Thus, the equation of motion of an antiQN inside the attracting body reads
\begin{equation}
    m_\text{QN}(t)\ddot{\mathbf{r}} = -\frac{G m_\text{QN}(t)M_\text{int}(r)\mathbf r}{r^3}
    -\frac12 \sigma_\text{col}\rho(r)v{\mathbf v}\,.
\label{A1}
\end{equation}
Note that there is also a linear friction term in velocity in this equation due to the recoil effect of the annihilation products. Specifically, nucleon annihilation predominantly occurs on the front side of the antiQN, and some annihilation products escape from this side. This anisotropy in the velocity distribution of the annihilation products can contribute to the friction force. However, for $\xi < 3 \times 10^{-4}$, nucleon annihilation mainly transpires inside the antiQN, and the momentum of the annihilation pions is absorbed within it. In this scenario, the recoil effect is negligible. Therefore, the recoil effect should not significantly influence our conclusions.
 
In Eq.~\eqref{A1} we ignored the Lorentz force acting on a charged particle in a magnetic field. Indeed, it is possible to show that this force is negligible as compared with the gravitational attraction and collisional friction because the magnetic fields of the Sun and Earth are relatively weak. Although the electric charge number of an antiQN in a dense medium like the Earth or Sun may reach $10^{5}$ \cite{FS0}, the corresponding gyroradius is of order $10^{15}$\,km.

AntiQNs lose their mass upon collisions with the visible matter. The mass loss equation is  specified by the annihilation cross section, $\sigma_\text{ann} = \kappa \pi R_\text{QN}^2 = \kappa\pi (1\,\text{fm}^2)A(t)^{2/3}$, $\kappa\simeq 0.1$ is the annihilation suppression coefficient,
\begin{equation}
    \frac{d m_\text{QN}(t)}{dt} = -\sigma_\text{ann} \rho v\,.
    \label{A2}
\end{equation}

It is easy to solve equations (\ref{A1}) and (\ref{A2}) numerically with initial condition for the antiQN velocity at large distance of about $250$\,km/s and different impact parameters. The solutions are qualitatively different for various values of the mass number $A$. Below we discuss these solutions separately for the Sun and the Earth.

\subsection{Annihilation of antiQNs in the Sun}

The gravitational field of the Sun is relatively strong, so that the antiQN speed near the Sun's surface becomes $v_0\simeq 650$\,km/s if it was 250 km/s at large distance from the Sun. We solve Eqs.~(\ref{A1}) and (\ref{A2}) numerically using the Sun density model from the NASA web site \cite{NASA},
\begin{align}
    &\rho(r) = \rho_0 \bigg[1-\frac{889}{155}\frac{r}{R_\odot}
    + \frac{1844}{155}\left(
    \frac{r}{R_\odot}
    \right)^2\nonumber\\&
    -\frac{1630}{155}
    \left(
    \frac{r}{R_\odot}
    \right)^3
    +\frac{519}{155}\left(
    \frac{r}{R_\odot}
    \right)^4\bigg]\,, \ \ 
    (r\leq R_\odot)
\end{align}
with $\rho_0 = 155$\,g/cm$^3$ the matter density in the Sun's core. We consider antiQNs with the initial velocity $v_0$ near the Sun's surface and impact parameter\footnote{Note that here $b$ is the impact parameter in the vicinity of the Sun which is smaller than the traditionally used impact parameter at large distance from gravitating body.} $b$ ranging from 0 to $R_\odot$. With these initial parameters, we find the following results: antiQNs with mass number $A\leq 10^{30}$ stop and completely annihilate inside the Sun for any impact parameter $b$, $0\leq b< 0.95R_\odot$. 

\begin{figure}
    \centering
    \includegraphics[width=0.7\linewidth]{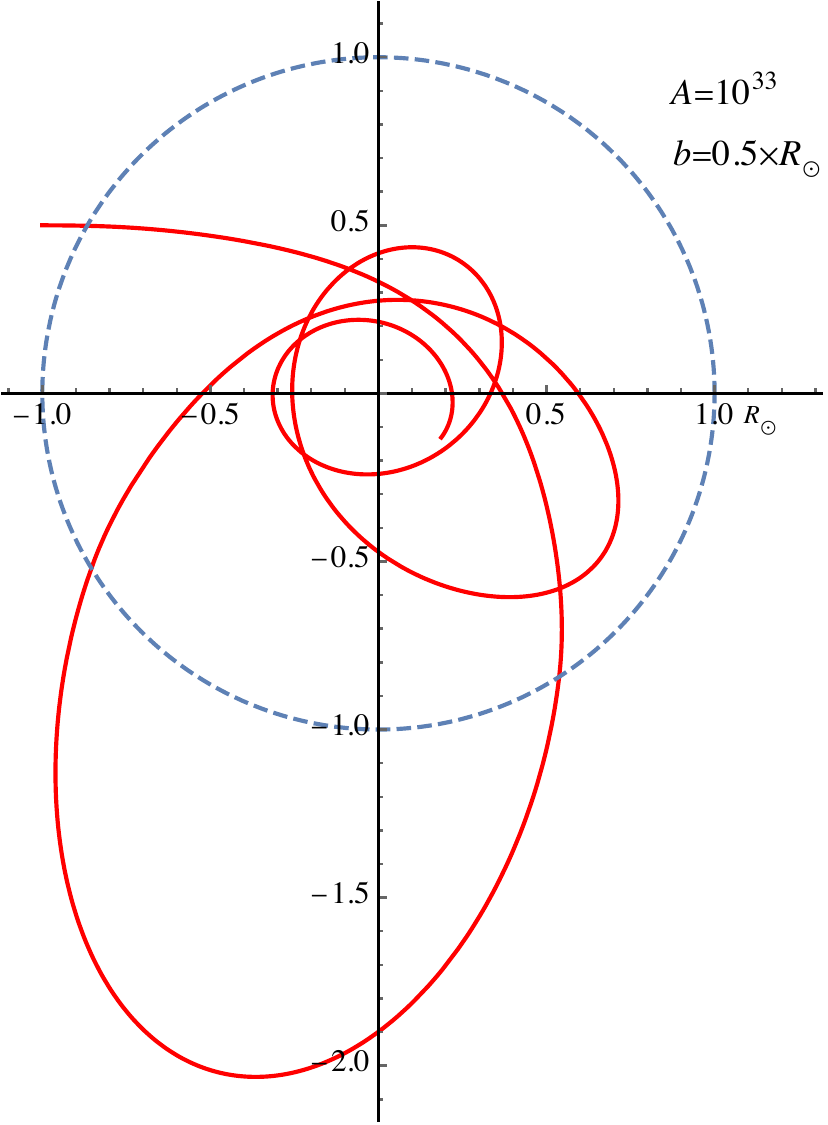}
    \caption{Sample trajectory (red) of an antiQN with $A=10^{33}$ annihilating inside the Sun (dashed circle). AntiQN entry point is on the left of the plot with impact parameter $b=0.5 R_\odot$. AntiQN possesses enough kinetic energy to traverse the Sun and exit, but it experience a significant energy loss, so that its exit velocity is less than the escape velocity at the Sun's surface.}
    \label{fig:Trajectory}
\end{figure}

AntiQNs with mass number $10^{31}\leq A \leq 10^{37}$ may traverse the Sun losing a fraction of their mass and exit the Sun with speed $v_\mathrm{exit}$. However, if $v_\mathrm{exit}<v_\mathrm{esc}=617$\,km/s, these DM particles are captured by the gravitational field of the Sun, and they fully annihilate upon subsequent passes through the Sun. A typical trajectory of such an antiQN is shown in Fig.~\ref{fig:Trajectory}. If $v_\mathrm{exit}>v_\mathrm{esc}$, antiQN escapes from the Solar system losing just a small fraction of its mass.

\begin{figure}
    \centering
    \includegraphics[width=0.9\linewidth]{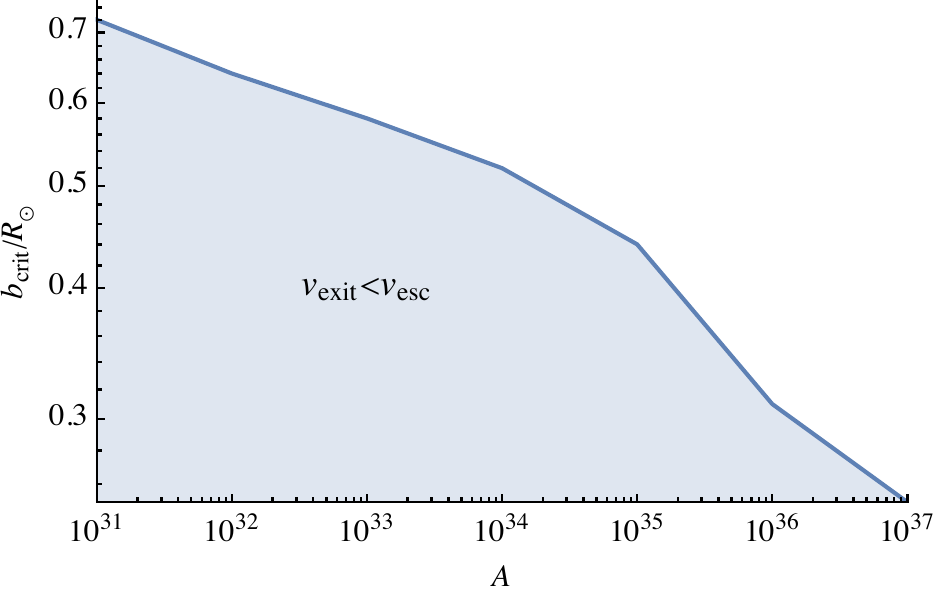}
    \caption{Dependence of the impact parameter $b_\mathrm{crit}$ on the mass number $A$. AniQNs impacting the Sun with $b<b_\mathrm{crit}$ exit the Sun with speed $v_\mathrm{exit}<v_\mathrm{esc}$ but remain bound by the Sun's gravitational field.}
    \label{fig:SunAnnihilation}
\end{figure}

For antiQNs with mass number $10^{31}\leq A \leq 10^{37}$ we define a critical value of the impact parameter $b_\mathrm{crit} = b_\mathrm{crit}(A)$ such that 
antiQNs with $b<b_\mathrm{crit}$ have $v_\mathrm{exit}<v_\mathrm{esc}$ and, thus, they fully annihilate inside the Sun. The plot of $b_\mathrm{crit}(A)$ is shown in Fig.~\ref{fig:SunAnnihilation}. In particular, antiQNs with $A\simeq 10^{33}$ have $b_\mathrm{crit}\simeq0.6R_\odot$ meaning that about 35\% of all antiQNs approaching the Sun from infinity are captured and eventually annihilate.

AntiQNs with $A>10^{37}$ are likely to escape from the Sun  losing less than 1\% of their mass even if they impact with $b=0$.

In this section, we have considered all antiQNs impacting the Sun with the same average speed of about 250 km/s at infinity. More generally, one could use a distribution of velocities of impacting DM particles. However, the present simplified consideration allows us to get an order-of-magnitude estimate of the effect of annihilation of antiQNs in the Sun depending on the mass number $A$. This accuracy is sufficient for the goals of this paper.
 
\subsection{Annihilation of antiQNs in the Earth}

As compared with the Sun, Earth's gravitational field is too weak to noticeably accelerate an antiQN approaching from infinity, and the size of the Earth is too small to fully stop antiQNs. Indeed, all antiQNs with $A>10^{24}$ pass through the Earth losing less than 10\% of their mass and escape from gravitational capturing. Their trajectories only slightly deviate from straight lines due to the gravitational attraction. AntiQNs with $A<10^{23}$ are likely to be stopped inside the Earth or gravitationally captured. However, the values $A<2\times 10^{24}$ are excluded by the present study, see Sec.~\ref{SecAnalysis}. This justifies the assumptions made in Sec.~\ref{SecEarth}.


\section{Photon pressure and annihilation efficiency}
\label{AppB}

In this Appendix, we discuss an effect which could potentially reduce the annihilation cross section of antiQNs with a dense matter like a rock on the Earth. Matter annihilation on antiQNs produces a photon flux which causes a photon pressure on atoms and molecules pushing them away from the trajectory of antiQNs. Let us estimate the maximum of the transferred momentum to an atom of matter situated on the antiQN's trajectory and compare the gained velocity with the antiQN typical speed $v_\mathrm{QN}\simeq 250$\,km/s.

\subsection{Contribution from $\gamma$-photons}

The main contribution to the photon pressure arises from $\gamma$-photons with energy $E_\gamma\sim 100-200$\,MeV originating from decaying $\pi^0$ mesons. These photons have an attenuation length $l\simeq 5-8$\,cm in a rock with average density $\rho=3-4$\,g/cm${}^3$. This means that about 63\% of the photon flux are absorbed within a cylinder of radius $l$ around the antiQN trajectory. An antiQN needs time $t=l/v_\mathrm{QN}$ to reach an atom at distance $l$ along its trajectory neglecting possible displacement of the atom due to the photon pressure. 

The nucleon number density of the matter is $n=\rho/m_p$, where $m_p$ is the proton rest mass. Matter annihilation rate on the antiQN with geometric cross section $\sigma = \pi\, \mathrm{fm}^2 A^{2/3}$ is $\Gamma = \kappa n v_\mathrm{QN} \sigma = \kappa v_\mathrm{QN} \pi\, \mathrm{fm}^2 A^{2/3}\rho/m_p$, with $\kappa$ the annihilation efficiency. Each matter-antimatter annihilation releases in total $E_0 = 2m_p c^2$ of energy, a fraction $\xi$ of which can be taken away by $\gamma$-photons. Thus, the energy emission rate is $F = \xi E_0 \Gamma = 2\xi c^2 \rho v_\mathrm{QN}\kappa\pi\,\mathrm{fm}^2 A^{2/3}$. The total momentum of these $\gamma$-photons emitted within the time interval $t=l/v_\mathrm{QN}$ is $p= F t /c=2\xi c \rho\kappa\pi\,\mathrm{fm}^2 A^{2/3} l$.

The total momentum of $\gamma$-photons is absorbed by all matter atoms within the  volume $V = \pi l^3$ which contains $N_\mathrm{at} = \rho V/m_\mathrm{at}$ atoms, where $m_\mathrm{at}$ is the mass of the atom. Thus, the average transferred momentum per atom in this volume is $\bar p = p/N_\mathrm{at} = 2\xi \kappa c A^{2/3} m_\mathrm{at} \mathrm{fm}^2/{l^2}$, and average atom speed is 
$\bar v = \bar p/m_\mathrm{at} = 2\xi \kappa c A^{2/3}\mathrm{fm}^2/{l^2}$. Imposing $\bar v<v_\mathrm{QN}$ \footnote{Note that this condition is valid when we take into account the component of the atomic velocity which is perpendicular to antiQN velocity, $v_\perp \sim v R_\mathrm{QN}/l$. This follows from the condition $v_\perp t \sim R_\mathrm{QN} v/v_\mathrm{QN} < R_\mathrm{QN}$.} we find the condition on the mass number of antiQN with insignificant $\gamma$-photon pressure on the surrounding matter,

\begin{equation}
    A\lesssim\left( \frac{ v_\mathrm{QN}l^2}{2c\xi \kappa \,\mathrm{fm}^2}
    \right)^{3/2}.
\end{equation}
For numerical estimates, we accept the following values of the parameters, $l=5$\,cm, $v_\mathrm{QN}=250$\,km/s, $\xi = 0.1$, $\kappa=0.1$. As a result, we find 
\begin{equation}
    A\lesssim10^{39}\,.
\end{equation}
Since in this paper we focus on antiQNs with $A<10^{33}$, we conclude that the $\gamma$-photon pressure on the surrounding matter is insignificant and may be ignored. Note that a contribution from electron-positron annihilation is a few orders in magnitude smaller and may be ignored as well.

\subsection{Contribution from thermal radiation}

If $\pi^0$ meson is absorbed deep inside antiQN core, there are no $\gamma$-photons from its decay. In this case, the coefficient $\xi$ is small, $\xi\ll1$, and the leading contribution to the photon pressure arises from thermal radiation produced by the position cloud of antiQN. 

As shown in Fig.~\ref{fig:Power}, 90\% of power of the thermal radiation is taken away by the photons with energy in the range from 50 to 600 keV. The attenuation length of such photons in a dense medium like a rock is $l=1-5$ cm. In our estimates below we will accept $l=1$\,cm to get an upper estimate of the effect of photon pressure. 

The contribution to the photon pressure from thermal radiation is similar to the considered above contribution from $\gamma$ photons with the only modification $\xi\to1-\xi$. Thus, an atom of the matter  may gain the average speed $\bar v = 2c\kappa(1-\xi) A^{2/3}\mathrm{fm}^2/l^2$. Imposing $\bar v<v_\mathrm{QN}$, we find the condition on the mass number of antiQN with insignificant contribution to the photon pressure from the thermal radiation:
\begin{equation}
    A\lesssim\left( \frac{ v_\mathrm{QN}l^2}{2c \kappa(1-\xi) \,\mathrm{fm}^2}
    \right)^{3/2}\simeq 10^{35}.
\end{equation}
This condition is satisfied for the considered in this paper antiQNs with $A<10^{33}$, and we conclude that photon pressure may be safely ignored.



\end{document}